\documentclass[12pt]{article}

\usepackage{amsmath}
\usepackage{amssymb}
\usepackage{amsfonts}
\usepackage{latexsym}
\usepackage{color}

\catcode `\@=11 \@addtoreset{equation}{section}

\catcode `\@=12

\newcommand{\be}{\begin{equation}}
\newcommand{\en}{\end{equation}}
\newcommand{\bea}{\begin{eqnarray}}
\newcommand{\ena}{\end{eqnarray}}
\newcommand{\beano}{\begin{eqnarray*}}
\newcommand{\enano}{\end{eqnarray*}}
\newcommand{\bee}{\begin{enumerate}}
\newcommand{\ene}{\end{enumerate}}

\newcommand{\mc}{\mathcal}

\newcommand{\D}{{\mc D}}

\newcommand{\Sc}{{\cal S}}
\newcommand{\G}{{\cal G}}

\newcommand{\F}{{\cal F}}

\newcommand{\1}{1 \!\! 1}

\newcommand{\Hil}{\mc H}

\catcode `\@=11 \@addtoreset{equation}{section}
\catcode `\@=12

\begin{document}

\thispagestyle{empty}

\vspace*{2cm}

\begin{center}
{\Large \bf Finite-dimensional  pseudo-bosons: a non-Hermitian version of the truncated harmonic oscillator}   \vspace{2cm}\\

{\large F. Bagarello}\\
 DEIM -Dipartimento di Energia, ingegneria dell' Informazione e modelli Matematici,
\\ Scuola Politecnica, Universit\`a di Palermo, I-90128  Palermo, Italy\\
and\\ INFN, Sezione di Napoli.\\
e-mail: fabio.bagarello@unipa.it

\end{center}

\vspace*{2cm}

\begin{abstract}
\noindent We propose a deformed version of the commutation rule introduced in 1967 by Buchdahl to describe a particular model of the truncated harmonic oscillator. The rule we consider is defined on a $N$-dimensional Hilbert space $\Hil_N$, and produces two biorhogonal bases of $\Hil_N$ which are eigenstates of the Hamiltonians $h=\frac{1}{2}(q^2+p^2)$, and of its adjoint $h^\dagger$. Here $q$ and $p$ are non-Hermitian operators obeying $[q,p]=i(\1-Nk)$, where $k$ is a suitable orthogonal projection operator. These eigenstates are connected by  ladder operators constructed out of $q$, $p$, $q^\dagger$ and $p^\dagger$. Some examples are discussed.

\end{abstract}

\vspace{2cm}



%{\bf PACS Numbers}:  .......

\vfill

%\pagenumbering{roman}

\newpage

\section{Introduction}\label{sectI}

Quantum mechanics is often thought to be naturally associated to self-adjoint (or Hermitian\footnote{We will use these two words as synonymous here.}) operators. In particular, the dynamics is deduced out of a self-adjoint Hamiltonian, and the {\em observables} of the system are almost always assumed to be self-adjoint as well.

In recent years, mainly since the seminal work by Bender and Boettcher, \cite{ben1}, it was understood that self-adjointness is not an essential requirement, since other operators exist, not self-adjoint, having purely real (and discrete) spectra. We refer to \cite{ben,mosta,bagbook} for some reviews on this alternative approach. What is interesting, from a mathematical point of view, is that orthonormal (o.n.) bases of eigenstates are replaced by biorthonormal sets which can be, or not, bases of the Hilbert space where the physical system lives. Also, different scalar products can play a role, and this different products produce different adjoints of the same operators. Moreover, the role of pseudospectra in connection with unbounded operators becomes relevant, \cite{petr}. Then, in a sense, loosing self-adjointness makes the mathematical structure reacher. Not only that: from a physical point of view the situation is also rather interesting since, for instance, some so-called PT-symmetric Hamiltonians can be naturally used to describe quantum systems with gain and loss phenomena, see \cite{gain1,gain2} and references therein.

In recent years, in connection with this kind of operators, we have developed a rather general formalism based on some suitable deformations of the canonical commutation and anti-commutation relations (CCR and CAR). These deformations produce what we have called $\D$-pseudo bosons and  pseudo-fermions. A rather complete review on both these topics can be found in \cite{baginbagbook}, to which we refer for several details and for some physical applications. Later on a similar framework was proposed for quons and for generalized Heisenberg algebra, \cite{bagquons,bagcurgaz}.

Here we consider a deformation of a different commutation rule, originally considered in \cite{buc}, and later analyzed in \cite{bagchi}, in connection with a truncated version of the harmonic oscillator. The operator $c$ considered in these papers obeys the following rule
\be
[c,c^\dagger]=\1-N\,K,
\label{11}\en
in which $N=2,3,4,\ldots$ is a fixed natural number, while $K$ is a self-adjoint projection operator, $K=K^2=K^\dagger$, satisfying the equality $Kc=0$. The presence of the term $N\,K$ in (\ref{11}) makes it possible to find a representation of $K$ and $c$ in terms of $N\times N$ matrices. In fact,  in absence of this term we would recover the CCR, which does not admit any finite-dimensional representation. Here, on the other hand, $K$, $c$ and $c^\dagger$ act on a $N$-th dimensional Hilbert space, which we call $\Hil_N$. In \cite{buc} it was shown that the matrices for $c$ and $c^\dagger$ are essentially the truncated versions of the analogous, infinite-dimensional, matrices for the bosonic annihilation and creation operators. In \cite{buc} it was also discussed how to construct an orthonormal (o.n.) basis of eigenvectors of the self-adjoint operator $H_0=\frac{1}{2}(Q_0^2+P_0^2)$, where $Q_0=\frac{c+c^\dagger}{\sqrt{2}}$ and $P_0=\frac{c-c^\dagger}{\sqrt{2}\,i}$ are the {\em truncated } position and momentum operators. These vectors turn out to be eigenvectors of both $H_0$ and $K$, and their explicit construction is strongly based on the fact that $H_0$ is a positive operator, other than being self-adjoint. This automatically imposes a lower bound on the possible eigenvalues of $H_0$, bound which was used in \cite{buc} to construct the set of eigenvectors. We will see that, in our extended case, positivity is apparently lost, so that we cannot adopt the same construction as in \cite{buc} for the  eigenvectors of our new Hamiltonian $h$, constructed in analogy with $H_0$. Moreover, since $h\neq h^\dagger$, it is natural to analyze also what happens for $h^\dagger$, and this will produce a biorthogonal set of eigenvectors of $h^\dagger$, see Section \ref{sectIII}, which is a basis for $\Hil_N$.

The article is organized as follows: in the next section we discuss our deformed version of the commutation rule (\ref{11}), and we construct a set of eigenvectors for the related truncated non self-adjoint harmonic oscillator, with Hamiltonian $h$, see above. We call the operators $a$ and $b$ appearing in this deformation {\em finite-dimensional pseudo-bosons} (FDPBs), since they can be seen as a truncated version of the $\D$-PBs considered in \cite{baginbagbook}. We show explicitly how our construction works for some fixed values of $N$, and then we generalize the procedure to generic $N$. Incidentally we will find that the procedure proposed here is much direct than that considered in \cite{buc}.  In Section \ref{sectIII} the biorthogonal set of eigenvectors of $h^\dagger$ is deduced. We also show how these FDPBs are related to the operators $c$ and $K$ in (\ref{11}). In Section \ref{sectIV} we discuss two examples, while our conclusions are given in Section \ref{sectV}.

\section{Deformed commutation rules}\label{sectII}

The main object of our research is the following deformed version of the commutation rule (\ref{11}):
\be
[a,b]=\1-Nk.
\label{21}\en
Here $N$ can be any fixed integer larger than 1, and $k$ is an orthogonal projector: $k=k^2=k^\dagger$. Extending what is done in \cite{buc} we also require that $ka=bk=0$. Moreover, $a$ and $b$ are not, in general, one the adjoint of the other: $b\neq a^\dagger$. This is, in a sense, close to what was done in \cite{pb1} first, and in \cite{pf1} later, for CCR and CAR, and, in fact, what we will show here, is that we recover the same {\em global} functional structure (raising and lowering relations, biorthogonal sets, non-Hermitian number-like operators,....) as in the cited papers, even if we work here in finite-dimensional Hilbert spaces of dimension not necessarily equal to 2, as we did in \cite{pf1}.

The first remark is that operators obeying the commutation rule in (\ref{21}) can also be represented as matrices acting on a $N$-dimensional Hilbert space $\Hil_N$. This can be easily seen as follows: let $S_0$ be an $(N-1)\times(N-1)$ invertible matrix, and let $s$ be a non-zero complex number. Then, if $S$ is the diagonal block matrix with blocks $S_0$ and $s$, $S^{-1}$ exists (but, in general, $S^{-1}\neq S^\dagger$) and, since (\ref{11}) is implemented in $\Hil_N$, we can easily define three new  matrices $a=ScS^{-1}$, $b=Sc^\dagger S^{-1}$ and $k=SKS^{-1}$. These operators, since $K$ commutes with $S^\dagger S$,satisfy (\ref{21}), as well as the equalities  $k=k^2=k^\dagger$ and $ka=bk=0$. So we see that, at least in this situation, (\ref{21}) can be represented in $\Hil_N$. Of course, other (higher-dimensional) representations could also exist. However, from now on, $a$, $b$ and $k$ will be considered as operators on $\Hil_N$.

We start our analysis by introducing two (non-Hermitian) position and momentum-like operators:
$$
q=\frac{a+b}{\sqrt{2}},\qquad p=\frac{a-b}{\sqrt{2}\,i},
$$
so that $a=\frac{q+ip}{\sqrt{2}}$ and $b=\frac{q-ip}{\sqrt{2}}$. As in \cite{buc}, we introduce the operator $h=\frac{1}{2}(p^2+q^2)$. Despite of its expression, and of what happens in \cite{buc}, $h$ is not Hermitian ($h\neq h^\dagger$). Moreover, it is  not even manifestly  positive ($h\ngeq0$), due to the fact that both $q$ and $p$ are not Hermitian. Nevertheless, we will show later in this section that the eigenvalues of $h$ are indeed strictly positive for all possible choices of $N$. After few computations it is easy to deduce the following equalities:
\be
[a,h]=a-\frac{1}{2}\,Nak,\qquad
[b,h]=-b+\frac{1}{2}\,Nkb,
\label{22}\en
as well as
\be
\left\{
    \begin{array}{ll}
h=ba+\frac{1}{2}(\1-Nk)=ab-\frac{1}{2}(\1-Nk),\\
\{a,b\}=2h,\\
kh=hk=-\frac{1}{2}(\1-N)k,
\end{array}
        \right.
\label{23}\en
which in particular imply that $[h,k]=0$.  Then we can look for common eigenstates of $h$ and $k$, which we call $\varphi_{h',k'}$:
\be
\left\{
    \begin{array}{ll}
h\varphi_{h',k'}=h'\varphi_{h',k'},\\
k\varphi_{h',k'}=k'\varphi_{h',k'}.
\end{array}
        \right.
\label{23a}\en
Of course, since $k=k^2$, $k'$ can only be 0 and 1. In particular, in analogy with what happens in \cite{buc}, the only (possibly) non zero vector $\varphi_{h',k'}$, when $k'=1$, is the vector with $h'=\frac{1}{2}(N-1)$, $\varphi_{\frac{1}{2}(N-1),1}$; all the other vectors, $\varphi_{h',1}$, if $h'\neq \frac{1}{2}(N-1)$, turn out to be zero. In general, the vectors $\varphi_{h',k'}$ are not mutually orthogonal in $h'$, since $h\neq h^\dagger$, while they are orthogonal in $k'$, since $k=k^\dagger$:
\be
\left<\varphi_{h',k'},\varphi_{h'',k''}\right>=\left<\varphi_{h',k'},\varphi_{h'',k'}\right>\,\delta_{k',k''}.
\label{24}\en

It is now possible to prove that, if $a\,\varphi_{h',k'}\neq0$, then this vector must be proportional to $\varphi_{h'-1+\frac{1}{2}Nk',0}$. This follows from the following facts: first, since $ka=0$, $k(a\varphi_{h',k'})=0$. Secondly, using (\ref{22}), we have
$$
h\left(a\varphi_{h',k'}\right)=\left([h,a]+ah\right)\varphi_{h',k'}=\left(h'-1+\frac{1}{2}Nk'\right)(a\varphi_{h',k'}).
$$
Hence our claim follows. In particular we have
\be
a\varphi_{h',0}=0 \quad \Leftrightarrow \quad h'=\frac{1}{2}.
\label{25}\en
In fact, let us assume that $\varphi_{h',0}\neq0$ but $a\varphi_{h',0}=0$. Then, using (\ref{23}), we have
$$
0=b\left(a\varphi_{h',0}\right)=\left(h-\frac{1}{2}(\1-Nk)\right)\varphi_{h',0}=\left(h'-\frac{1}{2}\right)\varphi_{h',0},
$$
so that $h'=\frac{1}{2}$. The proof of the converse implication, i.e. that $a\varphi_{\frac{1}{2},0}=0$, needs to be postponed but it is essentially based on the fact that $\Hil_N$ has dimension $N$. In fact, we will see that acting with $a$ and $b$ on vectors of the form $\varphi_{h',k'}$ we can produce $N$ linearly independent (l.i.) vectors, including $\varphi_{\frac{1}{2},0}$. Their linear independence is due to the fact that they correspond to different, strictly positive, values of $h'$ (so, even if they are not orthogonal, they are still l.i.), or to different values of $k'$ (so they are orthogonal and, therefore, l.i., too). Then, if $a\varphi_{\frac{1}{2},0}$ is different from zero, it would be proportional to $\varphi_{-\frac{1}{2},0}$. This vector, being eigenstate of $h$ with eigenvalue $h'=-\frac{1}{2}$ different from the other ones (see below), would be the $N+1$-th l.i. vector in a space with dimension $N$. This is clearly impossible. Hence (\ref{25}) follows. Notice that, in particular, this also implies that $h$ admits only strictly positive eigenvalues, even in absence of an manifest positivity, which was used in \cite{buc} to deduce the analogous of (\ref{25}).

After showing that $a$ annihilates $\varphi_{\frac{1}{2},0}$, we need now to show that $b$ annihilates the vector $\varphi_{\frac{1}{2}(N-1),1}$:
\be
b\varphi_{\frac{1}{2}(N-1),1}=0.
\label{26}\en
 To check this, we start observing that, if $kb\varphi_{h',k'}\neq0$, then it must be proportional to $\varphi_{h'+1-\frac{1}{2}N,1}$. First of all, since $k^2=k$, it is clear that in this case $kb\varphi_{h',k'}$ must
an eigenstate of $k$ with eigenvalue 1. Now, using (\ref{22}) we find first that
$$
hb\varphi_{h',k'}=\left(bh+b-\frac{1}{2}\,Nkb\right)\varphi_{h',k'}=\left(h'+1-\frac{1}{2}\,Nk\right)b\varphi_{h',k'},
$$
which, when left-multiplied by $k$, produces
$$
h\left(kb\varphi_{h',k'}\right)=\left(h'+1-\frac{1}{2}\,N\right)\left(kb\varphi_{h',k'}\right).
$$
Then, as stated,  if $kb\varphi_{h',k'}\neq0$, it must be proportional to $\varphi_{h'+1-\frac{1}{2}N,1}$. It is now possible to see that the only possibility for having $kb\varphi_{h',k'}\neq0$ is that $h'=N-\frac{3}{2}$. In other words: if $h'\neq N-\frac{3}{2}$ then $kb\varphi_{h',k'}=0$, independently of $k'$.

To prove this claim we use (\ref{22}) and the equality $kh=-\frac{1}{2}(\1-N)k$ in (\ref{23}). Few algebraic manipulations produce now the equality $kbh=\left(N-\frac{3}{2}\right)kb$. Therefore $\left(h'-\left(N-\frac{3}{2}\right)\right)kb\varphi_{h',k'}=0$, which is only possible, if $kb\varphi_{h',k'}\neq0$, when $h'=N-\frac{3}{2}$. Then, if $h'\neq N-\frac{3}{2}$, $kb\varphi_{h',k'}$ must be zero. If we now compute  $kb\varphi_{N-\frac{3}{2},k'}$ we find, because of what deduced before, that this is proportional to $\varphi_{\frac{1}{2}(N-1),1}$. Notice that, in principle, $kb\varphi_{N-\frac{3}{2},k'}$ could still be zero. However, inspired by the results in \cite{buc}, we will assume that this is not so, and check this assumption in explicit examples.

Formula (\ref{26}) is now a simple consequence of the fact that $bk=0$. Indeed we have, for what we have deduced so far, that $b\varphi_{\frac{1}{2}(N-1),1}$ must be proportional to $b\left(kb\varphi_{N-\frac{3}{2},k'}\right)$, which is the zero vector since $bk=0$. Summarizing we have two different vectors, $\varphi_{\frac{1}{2},0}$ and $\varphi_{\frac{1}{2}(N-1),1}$, which are annihilated respectively by $a$ and $b$. This is very close to what happens in \cite{buc}, where we have also two vectors annihilated by the operator $c$ in (\ref{11}), and by its hermitian conjugate $c^\dagger$. A similar feature is observed in ordinary CAR, where the lowering operator annihilates the vacuum and its adjoint annihilates the upper lever. Moreover, in \cite{buc}, it is shown that $c$ behaves as a sort of lowering operator, while $c^\dagger$ behaves as a raising operator. We expect that a similar behavior can be deduced here for $a$ and $b$, and this is in-fact what we will see now.

\subsection{Two preliminary examples}

Before discussing the general case (i.e. generic $N>2$), we briefly discuss   how the construction works when $N=2$ and when $N=3$. It is worth stressing that our construction is significantly different from the one proposed in \cite{buc}, because of the many properties related to the particular structure arising from (\ref{11}), which are lost here.

\vspace{2mm}

If $N=2$ the commutation rule in (\ref{21}) become $[a,b]=\1-2k$. The Hilbert space $\Hil_2$ is two-dimensional, and the two vectors in (\ref{25}) and (\ref{26}), $\varphi_{\frac{1}{2},0}$ and $\varphi_{\frac{1}{2},1}$, turn out to be orthogonal: $\left<\varphi_{\frac{1}{2},0},\varphi_{\frac{1}{2},1}\right>=0$, since they correspond to different eigenvalues of the Hermitian operator $k$. Then $\F_\varphi^{(2)}=\{\varphi_{\frac{1}{2},0},\varphi_{\frac{1}{2},1}\}$ is an o.n. basis for $\Hil_2$ (assuming a {\em good} normalization). Of course, the eigenvalue $\frac{1}{2}$ of the Hamiltonian $h$ is degenerate.
As discussed before, $a\varphi_{\frac{1}{2},0}$ must be zero. Otherwise, it would be proportional to $\varphi_{-\frac{1}{2},0}$, so that the vectors $\varphi_{\frac{1}{2},0}$, $\varphi_{\frac{1}{2},1}$ and $a\varphi_{\frac{1}{2},0}$ would be l.i., in contrast with the fact that $\dim(\Hil_2)=2$. Hence $a\varphi_{\frac{1}{2},0}=0$.

We have shown in (\ref{26}) that $b\varphi_{\frac{1}{2},1}=0$. So, in order to fully understand the situation, in this simple case we still need to compute $a\varphi_{\frac{1}{2},1}$ and $b\varphi_{\frac{1}{2},0}$. 
We will now prove that a non zero constant $\nu_{\frac{1}{2},0}$ exists such that
\be
a\varphi_{\frac{1}{2},1}=\nu_{\frac{1}{2},1}\,\varphi_{\frac{1}{2},0}, \qquad \mbox{and}\qquad b\varphi_{\frac{1}{2},0}=\nu_{\frac{1}{2},1}^{-1}\varphi_{\frac{1}{2},1}.
\label{27}\en
In fact, since $ha=ah-a+ak$, we see first that $ha\varphi_{\frac{1}{2},1}=(ah-a+ak)\varphi_{\frac{1}{2},1}=\frac{1}{2}a\varphi_{\frac{1}{2},1}$.
Moreover $k\left(a\varphi_{\frac{1}{2},1}\right)=0$. Then, a non zero constant  $\nu_{\frac{1}{2},1}$ should exist such that  $a\varphi_{\frac{1}{2},1}=\nu_{\frac{1}{2},1}\varphi_{\frac{1}{2},0}$. As for the second relation in (\ref{27}), this can be deduced by applying now the operator $b$ from the left to the first one and using the equality $[a,b]=\1-2k$.

%In Figure \ref{fig1} we show how $a$ and $b$ act on the two vectors of the basis of $\Hil_2$: $a$ behaves as a lowering operator, and destroys $\varphi_{\frac{1}{2},0}$. On the other hand, $b$ can be considered as a raising operator, and destroys $\varphi_{\frac{1}{2},1}$.
%
%
%\begin{figure}%[ht]
%\begin{center}
%
%\begin{picture}(250,115)
%
%\put(70,127){\line(1,0){100}} \put(72,127){\vector(-1,0){2}}\put(120,145){\makebox(0,0){$b$}}
%\put(70,135){\line(1,0){100}} \put(170,135){\vector(1,0){2}}\put(120,120){\makebox(0,0){$a$}}
%
%
%\multiput(50,120)(0,-5){10}{\line(0,-1){2}}\put(42,95){\makebox(0,0){$a$}}
%
%\multiput(180,120)(0,-5){10}{\line(0,-1){2}}\put(188,95){\makebox(0,0){$b$}}
%
%
%\put(55,130){\makebox(0,0){$\varphi_{\frac{1}{2},0}$}}\put(50,70){\vector(0,-1){2}}
%\put(185,130){\makebox(0,0){$\varphi_{\frac{1}{2},1}$}}\put(180,70){\vector(0,-1){2}}
%
%\put(50,60){\makebox(0,0){$0$}}\put(180,60){\makebox(0,0){$0$}}
%
%\end{picture}
%
%\end{center}
%
%\vspace{-1.5cm}
%
%
%\caption{{\protect\footnotesize The ladder operators and their action in $\Hil_2$.}}
%\label{fig1}
%\end{figure}

\vspace{2mm}

The case $N=3$ is surely more interesting because, as we will see, the vectors $\varphi_{h',k'}$ do not form an o.n. basis for $\Hil_3$.

In this case the commutation rule in (\ref{21}) becomes $[a,b]=\1-3k$ and, since $\frac{1}{2}(N-1)=1$, our {\em extreme} vectors in (\ref{25}) and (\ref{26}) are $\varphi_{\frac{1}{2},0}$ and $\varphi_{1,1}$. Since they correspond to  different eigenvalues of the Hermitian operator $k$, they are orthogonal: $\left<\varphi_{\frac{1}{2},0},\varphi_{1,1}\right>=0$. But they are just two vectors in a three-dimensional space. Therefore, they cannot be a basis for $\Hil_3$. A third vector, l.i. with respect to the these  two, can be easily constructed. To do this, let us consider the vector $a\varphi_{1,1}$. Using (\ref{22}) and the fact that $ka=0$ we deduce that
$$
h\left(a\varphi_{1,1}\right)=\frac{3}{2}\left(a\varphi_{1,1}\right), \qquad k\left(a\varphi_{1,1}\right)=0.
$$
These imply that, if $a\,\varphi_{1,1}\neq0$, then it is an eigenvector of $h$, with eigenvalue $\frac{3}{2}$, and of $k$, with eigenvalue $0$. Hence we can introduce a vector, $\varphi_{\frac{3}{2},0}$, and a non-zero (complex) number $\nu_{1,1}$, such that $a\,\varphi_{1,1}=\nu_{1,1}\varphi_{\frac{3}{2},0}$. Let then consider the set $\F_\varphi^{(3)}=\{\varphi_{\frac{1}{2},0},\varphi_{\frac{3}{2},0},\varphi_{1,1}\}$. Its vectors are l.i., since they are orthogonal or they correspond to different eigenvalues of $h$. Hence $\F_\varphi^{(3)}$ is a basis for $\Hil_3$. Furthermore, using the commutation rule between $a$ and $b$, it follows that $b\,\varphi_{\frac{3}{2},0}=2\nu_{1,1}^{-1}\varphi_{1,1}$. In a similar way we can also prove that
$$
a\,\varphi_{\frac{3}{2},0}=\nu_{\frac{3}{2},0}\,\varphi_{\frac{1}{2},0},\qquad \mbox{ and } \qquad b\,\varphi_{\frac{1}{2},0}=\nu_{\frac{3}{2},0}^{-1}\,\varphi_{\frac{3}{2},0},
$$
for some non zero $\nu_{\frac{3}{2},0}$. This is because  $a\,\varphi_{\frac{3}{2},0}$ turns out to be an eigenstate of $h$ and $k$, with eigenvalues $\frac{1}{2}$ and $0$ respectively. Hence the first equality above follows. The second is a consequence of this first and of the commutator $[a,b]=\1-3k$. We see that, with a slight abuse of language, $a$ behaves as a lowering operator while $b$ behaves as a raising operator.

\subsection{Larger $N$}

%once again, we start with the two vectors $\varphi_{\frac{1}{2},0}$ and $\varphi_{\frac{1}{2}(N-1),1}=\varphi_{\frac{3}{2},1}$ in (\ref{25}) and (\ref{26}), which are obviously orthogonal, and we look for two more l.i. vectors. For that, in analogy with the case $N=3$, we act on the upper vector $\varphi_{\frac{3}{2},1}$ with $a$, getting, if $a\varphi_{\frac{3}{2},1}\neq0$, an eigenstate of $h$ and $k$ with eigenvalues $\frac{5}{2}$ and $0$ respectively. Therefore $a\varphi_{\frac{3}{2},1}=\nu_{\frac{3}{2},1}\varphi_{\frac{5}{2},0}$, for some non zero $\nu_{\frac{3}{2},1}$.  Hence we have three l.i. vectors in $\Hil_4$. The fourth vector of the basis  $\F_\varphi^{(4)}$ of $\Hil_4$ can be constructed by acting again with $a$ on $\varphi_{\frac{5}{2},0}$. The vector $a\varphi_{\frac{5}{2},0}$, if it is non zero, is an eigenstate of $h$ and $k$ with eigenvalues $\frac{3}{2}$ and $0$ respectively. Therefore $a\varphi_{\frac{5}{2},0}=\nu_{\frac{5}{2},0}\varphi_{\frac{3}{2},0}$, for some non zero $\nu_{\frac{5}{2},0}$. The set $\F_\varphi^{(4)}=\{\varphi_{\frac{1}{2},0},\varphi_{\frac{3}{2},0},\varphi_{\frac{5}{2},0},\varphi_{\frac{3}{2},1}\}$ is the basis of $\Hil_4$ we were looking for: we have four vectors which are l.i. either because they correspond to different eigenvalues of $k$ (and in this case they are mutually orthogonal) or because they correspond to different eigenvalues of $h$.

\vspace{3mm}

\begin{figure}%[ht]
\begin{center}

\begin{picture}(450,145)

\put(5,127){\line(1,0){65}} \put(7,127){\vector(-1,0){2}}\put(40,145){\makebox(0,0){$b$}}
\put(5,135){\line(1,0){65}} \put(68,135){\vector(1,0){2}}\put(40,120){\makebox(0,0){$a$}}
\multiput(-10,120)(0,-5){10}{\line(0,-1){2}}\put(-15,95){\makebox(0,0){$a$}}\put(-10,70){\vector(0,-1){2}}
\put(-10,60){\makebox(0,0){$0$}}

\put(100,127){\line(1,0){65}} \put(103,127){\vector(-1,0){2}}\put(130,145){\makebox(0,0){$b$}}
\put(100,135){\line(1,0){65}} \put(163,135){\vector(1,0){2}}\put(130,120){\makebox(0,0){$a$}}

\put(310,127){\line(1,0){65}} \put(313,127){\vector(-1,0){2}}\put(340,145){\makebox(0,0){$b$}}
\put(310,135){\line(1,0){65}} \put(373,135){\vector(1,0){2}}\put(340,120){\makebox(0,0){$a$}}

\multiput(200,130)(3,0){20}{\line(2,0){1}}

%\multiput(150,120)(0,-5){10}{\line(0,-1){2}}\put(188,95){\makebox(0,0){$b$}}

\multiput(395,120)(0,-5){10}{\line(0,-1){2}}\put(400,95){\makebox(0,0){$b$}}\put(395,70){\vector(0,-1){2}}
\put(395,60){\makebox(0,0){$0$}}

\put(-10,130){\makebox(0,0){$\varphi_{\frac{1}{2},0}$}}
\put(85,130){\makebox(0,0){$\varphi_{\frac{3}{2},0}$}}
\put(180,130){\makebox(0,0){$\varphi_{\frac{5}{2},0}$}}
\put(285,130){\makebox(0,0){$\varphi_{\frac{N}{2}-3,0}$}}
\put(405,130){\makebox(0,0){$\varphi_{\frac{1}{2}(N-1),1}$}}

\end{picture}

\end{center}

\vspace{-1.5cm}

\caption{{\protect\footnotesize The ladder operators and their action in $\Hil_N$.}}
\label{fig3}
\end{figure}
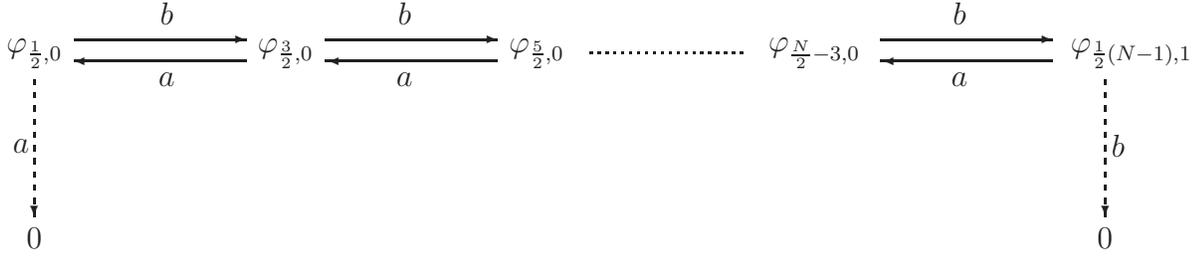

The situation is just a little more complicated if we take now $N\geq4$: 
 we have two vectors $\varphi_{\frac{1}{2},0}$ and $\varphi_{\frac{1}{2}(N-1),1}$ which are mutually orthogonal in $\Hil_N$. A third l.i. vector can be defined via the action of $a$ on $\varphi_{\frac{1}{2}(N-1),1}$. In fact $a\varphi_{\frac{1}{2}(N-1),1}$, if it is non-zero, is an eigenstate of $h$ and $k$ with eigenvalues $N-\frac{3}{2}$ and $0$ respectively:
$$
h\left(a\varphi_{\frac{1}{2}(N-1),1}\right)=\left(N-\frac{3}{2}\right)\left(a\varphi_{\frac{1}{2}(N-1),1}\right), \qquad\mbox{  and }\qquad k\left(a\varphi_{\frac{1}{2}(N-1),1}\right)=0.
$$
Then, a non zero constant $\nu_{\frac{1}{2}(N-1),1}$ exists such that $a\varphi_{\frac{1}{2}(N-1),1}=\nu_{\frac{1}{2}(N-1),1}\varphi_{N-\frac{3}{2},0}$. Now, since $N\geq4$, $N-\frac{3}{2}\neq \frac{1}{2}$. Therefore $\{\varphi_{\frac{1}{2},0},\varphi_{N-\frac{3}{2},0},\varphi_{\frac{1}{2}(N-1),1}\}$ are l.i., for the usual reasons, but they cannot form a basis for $\Hil_N$. Acting with $b$ on  $\varphi_{N-\frac{3}{2},0}$ we go back to $\varphi_{\frac{1}{2}(N-1),1}$. More in details, $b\varphi_{N-\frac{3}{2},0}=(N-1)\nu_{\frac{1}{2}(N-1),1}^{-1}\varphi_{\frac{1}{2}(N-1),1}$. If we go further, considering now $a\varphi_{N-\frac{3}{2},0}$, we find that this vector satisfies the following eigenvalue equations:
$$
h\left(a\varphi_{N-\frac{3}{2},0}\right)=\left(N-\frac{5}{2}\right)\left(a\varphi_{N-\frac{3}{2},0}\right), \qquad\mbox{  and }\qquad k\left(a\varphi_{N-\frac{3}{2},0}\right)=0.
$$
Therefore, if $a\varphi_{N-\frac{3}{2},0}\neq0$, a non zero  $\nu_{N-\frac{3}{2},0}$ exists such that $a\varphi_{N-\frac{3}{2},0}=\nu_{N-\frac{3}{2},0}\varphi_{N-\frac{5}{2},0}$. Also, $b\varphi_{N-\frac{5}{2},0}=(N-2)\nu_{N-\frac{3}{2},0}^{-1}\varphi_{N-\frac{3}{2},0}$. Of course, if $N>4$, $N-\frac{5}{2}\neq \frac{1}{2}$. Therefore $\{\varphi_{\frac{1}{2},0},\varphi_{N-\frac{5}{2},0},\varphi_{N-\frac{3}{2},0},\varphi_{\frac{1}{2}(N-1),1}\}$ are still l.i., but, again, they cannot be a basis for $\Hil_N$. So we consider $a\varphi_{N-\frac{5}{2},0}$, and the usual arguments show that this is proportional to $\varphi_{N-\frac{7}{2},0}$. Hence we have two possibilities: either $N-\frac{9}{2}=\frac{1}{2}$, which means that $N=5$, or, when  $N-\frac{9}{2}>\frac{1}{2}$, $N>5$. In the first case $\F_\varphi^{(5)}=\{\varphi_{\frac{1}{2},0},\varphi_{N-\frac{7}{2},0},\varphi_{N-\frac{5}{2},0},
\varphi_{N-\frac{3}{2},0},\varphi_{\frac{1}{2}(N-1),1}\}$ are five l.i. vectors in $\Hil_5$, so they do form a basis. In fact, acting again with $a$ on $\varphi_{N-\frac{7}{2},0}$ would produce a vector proportional to $\varphi_{\frac{1}{2},0}$, which is clearly not l.i. with respect to the ones in $\F_\varphi^{(5)}$. On the other hand, if $N>5$, we can continue our lowering procedure, acting with $a$ on $\varphi_{N-\frac{7}{2},0}$ and getting, this time, a vector which is surely not proportional to $\varphi_{\frac{1}{2},0}$. Of course, while $a$ acts as a lowering operator, $b$ behaves as a raising operator. However, due to the fact that $\Hil_N$ is finite-dimensional, $b$ annihilates $\varphi_{\frac{1}{2}(N-1),1}$, as we have seen. The situation is shown in Figure \ref{fig3}.

\vspace*{2mm}

{\bf Remarks:--} (1) We first observe that the values of the $\nu_{h',k'}$ in our construction are unfixed. The reason is that they will be only (partially) fixed by the biorthogonality condition discussed in the next section.

(2) The vectors $\varphi_{h',k'}$ are eigenstates not only of $h$ and $k$, but also of the operators $\hat M=ab$ and $\hat N=ba$. In fact we find:
\be
\left\{
    \begin{array}{ll}
\hat N\varphi_{\frac{1}{2}(N-1),1}=(N-1)\varphi_{\frac{1}{2}(N-1),1},\\
\hat N\varphi_{N-\frac{3}{2},0}=(N-2)\varphi_{N-\frac{3}{2},0},\\
\hat N\varphi_{N-\frac{5}{2},0}=(N-3)\varphi_{N-\frac{5}{2},0},\\
\ldots\\
\ldots\\
\hat N\varphi_{\frac{1}{2},0}=0,
\end{array}
        \right.
\label{28}\en
while
\be
\left\{
    \begin{array}{ll}
\hat M\varphi_{\frac{1}{2}(N-1),1}=0,\\
\hat M\varphi_{N-\frac{3}{2},0}=(N-1)\varphi_{N-\frac{3}{2},0},\\
\hat M\varphi_{N-\frac{5}{2},0}=(N-2)\varphi_{N-\frac{5}{2},0},\\
\ldots\\
\ldots\\
\hat M\varphi_{\frac{1}{2},0}=\varphi_{\frac{1}{2},0}.
\end{array}
        \right.
\label{29}\en
This is not surprising since one can check that $[\hat N,h]=[\hat M,h]=[\hat N,k]=[\hat M,k]=0$. In fact, recalling that $ka=bk=0$, we see that $[k,\hat M]=[k,ab]=0$. Moreover, since $ba=ab+Nk-\1$, we deduce that $[k,\hat N]=[k,ba]=[k,ab+Nk-\1]=0$. The fact that we also have $[\hat N,h]=[\hat M,h]=0$ follows now from the first line in (\ref{23}) which implies that $\hat M=h+\frac{1}{2}(\1-Nk)$ and $\hat N=h-\frac{1}{2}(\1-Nk)$, recalling further that $[h,k]=0$.

\section{The biorthogonal set}\label{sectIII}

In the literature on non self-adjoint Hamiltonians it is widely discussed how biorthogonal sets play an essential role in the description of the physical system $\Sc$: the eigenstates of the Hamiltonian $H_\Sc$ of $\Sc$ are not orthogonal, while they are biorthogonal to the elements of the set of eigenstates of $H_\Sc^\dagger$. For this reason, recalling that the operator $h=\frac{1}{2}(p^2+q^2)$ introduced in Section \ref{sectII} is not Hermitian, $h\neq h^\dagger$, it is natural to consider the problem of the diagonalization of $h^\dagger$. For that we take the adjoint of (\ref{21}) and we observe that, introducing $A=b^\dagger$ and $B=a^\dagger$, and recalling that $k=k^\dagger$, we get
\be
[A,B]=\1-Nk.
\label{31}\en
Then the pair $(A,B)$ satisfies the same commutation rule as the original pair $(a,b)$. Now, calling $Q=\frac{A+B}{\sqrt{2}}$, $P=\frac{A-B}{\sqrt{2}\,i}$ and $H=\frac{1}{2}(P^2+Q^2)$, it follows that $Q=q^\dagger$, $P=p^\dagger$ and $H=h^\dagger$. The analogous of formulas (\ref{22}) and (\ref{23}) can be deduced for these operators. For instance $\{A,B\}=2H$, $BA=H-\frac{1}{2}(\1-Nk)$, $AB=H+\frac{1}{2}(\1-Nk)$, $[H,k]=0$ and so on. Moreover, we also have $Bk=kA=0$. We call now $\psi_{h',k'}$ the common eigenstates of $H$ and $k$:
\be
\left\{
    \begin{array}{ll}
H\psi_{h',k'}=h'\psi_{h',k'},\\
k\psi_{h',k'}=k'\psi_{h',k'}.
\end{array}
        \right.
\label{32}\en
Then, standard arguments show that the sets $\F_\varphi=\{\varphi_{h',k'}\}$ and $\F_\psi=\{\psi_{h',k'}\}$ are biorthogonal and, if we choose properly the normalization of the vectors, they are also biorthonormal:
\be
\left<\psi_{h',k'},\varphi_{h'',k''}\right>=\delta_{h',h''}\delta_{k',k''}.
\label{33}\en

The construction of the set $\F_\psi^{(N)}$ reflects that of $\F_\varphi^{(N)}$ for the different values of $N$. For instance, if $N=2$, we only have two vectors $\F_\psi^{(2)}=\{\psi_{\frac{1}{2},0},\psi_{\frac{1}{2},1}\}$ and they satisfy the following equations:
$$
A\psi_{\frac{1}{2},0}=B\psi_{\frac{1}{2},1}=0, \qquad \mbox{and }\quad A\psi_{\frac{1}{2},1}=\mu_{\frac{1}{2},1}\psi_{\frac{1}{2},0},\quad B\psi_{\frac{1}{2},0}=\mu_{\frac{1}{2},1}^{-1}\psi_{\frac{1}{2},1},
$$
for some non-zero $\mu_{\frac{1}{2},1}$. This has to be related to $\nu_{\frac{1}{2},1}$ if we want $\F_\varphi^{(2)}$ and $\F_\psi^{(2)}$ to be biorthonormal. In fact, while $\left<\psi_{\frac{1}{2},1},\varphi_{\frac{1}{2},0}\right>=
\left<\psi_{\frac{1}{2},0},\varphi_{\frac{1}{2},1}\right>=0,$
automatically, if we further choose the normalization of $\varphi_{\frac{1}{2},0}$ and $\psi_{\frac{1}{2},0}$ by requiring $
\left<\psi_{\frac{1}{2},0},\varphi_{\frac{1}{2},0}\right>=1,
$
then we also get
$
\left<\psi_{\frac{1}{2},1},\varphi_{\frac{1}{2},1}\right>=1
$
at least if \be \overline{\nu_{\frac{1}{2},1}}\,\mu_{\frac{1}{2},1}=1.\label{add1}\en A similar situation is recovered for larger values of $N$. For instance, if $N=3$, the set $\F_\psi^{(3)}=\{\psi_{\frac{1}{2},0},\psi_{\frac{3}{2},0},\psi_{1,1}\}$ obeys the following ladder rules, see Figure \ref{fig2}:
\be
 A\psi_{1,1}=\mu_{1,1}\psi_{\frac{3}{2},0},\quad A\psi_{\frac{3}{2},0}=\mu_{\frac{3}{2},0}\psi_{\frac{1}{2},0},\quad B\psi_{\frac{3}{2},0}=2\mu_{1,1}^{-1}\psi_{1,1}, \quad B\psi_{\frac{1}{2},0}=\mu_{\frac{3}{2},0}^{-1}\psi_{\frac{3}{2},0},
\label{add2}\en
for some suitable and non zero $\mu_{1,1}$ and $\mu_{\frac{3}{2},0}$, as well as $A\psi_{\frac{1}{2},0}=B\psi_{1,1}=0$. Using now (\ref{21}) and assuming that $\left<\psi_{\frac{1}{2},0},\varphi_{\frac{1}{2},0}\right>=1$ we recover biorthonormality of the sets $\F_\psi^{(3)}$ and $\F_\varphi^{(3)}$  if we require
\be\overline{\nu_{\frac{3}{2},0}}\,\mu_{\frac{3}{2},0}=\frac{1}{2}\,\overline{\nu_{1,1}}\,\mu_{1,1}=1.\label{34}\en
%Yet, for $N=4$, the set $\F_\psi^{(4)}=\{\psi_{\frac{1}{2},0},\psi_{\frac{3}{2},0},,\psi_{\frac{5}{2},0},\psi_{\frac{3}{2},1}\}$ is biorthonormal to $\F_\varphi^{(4)}$ if, as usual, $\left<\psi_{\frac{1}{2},0},\varphi_{\frac{1}{2},0}\right>=1$ and if
%\be
%\overline{\nu_{\frac{3}{2},0}}\,\mu_{\frac{3}{2},0}=\frac{1}{2}\,\overline{\nu_{\frac{5}{2},0}}\,\mu_{\frac{5}{2},0}=
%\frac{1}{3}\,\overline{\nu_{\frac{3}{2},1}}\,\mu_{\frac{3}{2},1}=1.
%\label{35}\en
It is straightforward to extend these results to larger $N$. In all cases, the following resolution of the identity in $\Hil_N$ is satisfied:
$$
\sum_{(h',k')\in\G_N}|\psi_{h',k'}\left>\right<\varphi_{h',k'}|=\sum_{(h',k')\in\G_N}|\varphi_{h',k'}\left>\right<\psi_{h',k'}|=\1,
$$
where we have used the Dirac bra-ket notation and where the sum is extended, for each fixed $N$, to all the possible pairs of $(h',k')$, see Section \ref{sectII}. We have called $\G_N$ this set.

\vspace{3mm}

The two biorthonormal sets $\F_\varphi^{(N)}$ and $\F_\psi^{(N)}$ can be used, together, to represent the operators $a$, $b$, and their adjoints, as a sum of rank-one operators. For instance, if $N=2$, we have
\be
a=\nu_{\frac{1}{2},1}\,|\varphi_{\frac{1}{2},0}\left>\right<\psi_{\frac{1}{2},1}|, \qquad b=\nu_{\frac{1}{2},1}^{-1}\,|\varphi_{\frac{1}{2},1}\left>\right<\psi_{\frac{1}{2},0}|,
\label{39}\en
whose adjoints are $a^\dagger=\overline{\nu_{\frac{1}{2},1}}\,|\psi_{\frac{1}{2},1}\left>\right<\varphi_{\frac{1}{2},0}|$ and $b^\dagger=\overline{\nu_{\frac{1}{2},1}^{-1}}\,|\psi_{\frac{1}{2},0}\left>\right<\varphi_{\frac{1}{2},1}|$. It is interesting to observe that the operators $a^\dagger$ and $b^\dagger$ given here coincide with $B$ and $A$ if condition (\ref{add1}) is satisfied.

In a similar way, if $N=3$, $a$ and $b$ can be written as
\be
a=\nu_{1,1}\,|\varphi_{\frac{3}{2},0}\left>\right<\psi_{1,1}|+\nu_{\frac{3}{2},0}\,
|\varphi_{\frac{1}{2},0}\left>\right<\psi_{\frac{3}{2},0}|, \qquad b=2\nu_{1,1}^{-1}\,|\varphi_{1,1}\left>\right<\psi_{\frac{3}{2},0}|+
\nu_{\frac{3}{2},0}^{-1}\,|\varphi_{\frac{3}{2},0}\left>\right<\psi_{\frac{1}{2},0}|,
\label{310}\en
and the adjoint $a^\dagger$ and $b^\dagger$ which we deduce out of these behaves as the operators $B$ and $A$ in (\ref{add2}) if conditions in (\ref{34}) are satisfied. The same is true for larger values of $N$. 

%For $N=4$ we have
%$$
%a=\nu_{\frac{3}{2},1}\,
%|\varphi_{\frac{5}{2},0}\left>\right<\psi_{\frac{3}{2},1}|+\nu_{\frac{5}{2},0}\,|\varphi_{\frac{3}{2},0}\left>
%\right<\psi_{\frac{5}{2},0}|+\nu_{\frac{3}{2},0}\,
%|\varphi_{\frac{1}{2},0}\left>\right<\psi_{\frac{3}{2},0}|,
%$$
%$$
%b=3\nu_{\frac{3}{2},1}^{-1}\,
%|\varphi_{\frac{3}{2},1}\left>\right<\psi_{\frac{5}{2},0}|+2\nu_{\frac{5}{2},0}^{-1}\,|\varphi_{\frac{5}{2},0}\left>
%\right<\psi_{\frac{3}{2},0}|+\nu_{\frac{3}{2},0}^{-1}\,
%|\varphi_{\frac{3}{2},0}\left>\right<\psi_{\frac{1}{2},0}|,
%$$
%and $a^\dagger$ and $b^\dagger$ follow from these ones. 

It turns out that  $a^\dagger$ and $b^\dagger$ are respectively equal to the ladder operators $B$ and $A$ for $\F_\psi^{(4)}$. This is a general characteristic of the construction: the same constraints on $\mu_{h',k'}$ and $\nu_{h',k'}$ which make of $\F_\varphi^{(N)}$ and $\F_\psi^{(N)}$ biorthonormal bases ensure that the adjoint of the representations of $a$ and $b$ coincide exactly with $B$ and $A$.

%\vspace{2mm}
%
%{\bf Remark:--} It should be mentioned that a similar procedure to construct ladder operators in presence of non Hermitian Hamiltonians was discussed in \cite{abg} in the context of pseudo-fermions. However, in \cite{abg} we were not able to produce, except for $N=2$, any useful commutation or anticommutation rule between these operators. Here, on the contrary, these rules are exactly our starting point. This probably makes the results deduced in this paper more interesting and more flexible for more concrete applications.

\subsection{Relation with (\ref{11})}\label{sectIII.1}

At the beginning of Section \ref{sectII} we have already discussed how (\ref{21}) can be obtained from (\ref{11}), by means of a similarity map. In this section we discuss the inverse construction, i.e. we show how, starting from (\ref{21}), it is possible to construct two operators, $c$ and $K$, which obey the commutation rule in (\ref{11}) and such that $Kc=0$. Our construction is similar, but not identical, to that proposed for pseudo-fermions, \cite{pf1}.

We start introducing the operators
\be
S_\varphi=\sum_{(h',k')\in\G_N}|\varphi_{h',k'}\left>\right<\varphi_{h',k'}|, \qquad S_\psi=\sum_{(h',k')\in\G_N}|\psi_{h',k'}\left>\right<\psi_{h',k'}|.
\label{36}\en
These are bounded, invertible, Hermitian and positive. Moreover, they are one the inverse of the other, $S_\varphi S_\psi=S_\psi S_\varphi=\1$, and satisfy the following:
\be
S_\varphi \psi_{h',k'}=\varphi_{h',k'}, \qquad S_\psi \varphi_{h',k'}=\psi_{h',k'},
\label{37}\en
for all $(h',k')\in\G_N$. $S_\psi$ admits an unique positive square root, which is also invertible. Hence we can define
\be
e_{h',k'}=S_\psi^{1/2}\varphi_{h',k'}, \qquad c=S_\psi^{1/2}aS_\psi^{-1/2},
\label{38}\en
for $(h',k')\in\G_N$.  Now, it is a simple computation to prove that the set $\F_e=\{e_{h',k'}\}$ is an o.n. basis for $\Hil_N$. It is also possible to check that $c^\dagger$, other than being equal to $S_\psi^{-1/2}a^\dagger S_\psi^{1/2}$, can also be written as $c^\dagger=S_\psi^{1/2}bS_\psi^{-1/2}$. In fact $S_\psi^{-1/2}a^\dagger S_\psi^{1/2}=S_\psi^{1/2}bS_\psi^{-1/2}$ if and only if $S_\psi b=a^\dagger S_\psi$, i.e. if $S_\psi$ intertwines between $b$ and $a^\dagger$. This equality can be easily deduced by considering the action of $S_\psi b$ and $a^\dagger S_\psi$ on the vectors of $\F_\varphi$. In fact, it turns out that $S_\psi b\varphi_{h',k'}=a^\dagger S_\psi\varphi_{h',k'}$ for all $(h',k')\in\G_N$, at least if we assume that $\mu_{h',k'}=\nu_{h',k'}$ for all $(h',k')\in\G_N$. This further constraint simplifies the conditions we have found before to guarantee the biorthonormality of the sets $\F_\varphi^{(N)}$ and $\F_\psi^{(N)}$ and in the analysis of the representation of the operators $a$, $a^\dagger$, $b$ and $b^\dagger$ in terms of these vectors. 

Now, since $c=S_\psi^{1/2}aS_\psi^{-1/2}$ and $c^\dagger=S_\psi^{1/2}bS_\psi^{-1/2}$, it turns out that $[c,c^\dagger]=\1-NK$, where $K$ is defined as $K=S_\psi^{1/2}kS_\psi^{-1/2}$. Of course $K^2=K$ and $Kc=0$. Moreover $K=K^\dagger$ if and only if $k$ commutes with $S_\psi$, which is true:
$
S_\psi k\varphi_{h',k'}=k'S_\psi \varphi_{h',k'}=k'\psi_{h',k'}=k\psi_{h',k'}=kS_\psi\varphi_{h',k'}.
$
Hence, since $\Hil_N$ has finite dimension, $[k,S_\psi]=0$.

\vspace{2mm}

Summarizing we can say that it is possible to deform (\ref{11}) to get (\ref{21}), but we can also go the other way around, i.e. we can consider (\ref{21}) as our starting point, and use the eigenvectors of $h$, $h^\dagger$ and $k$ constructed out of $a$ and $b$ to define new operators satisfying (\ref{11}).

\section{Examples}\label{sectIV}

In this section we consider a pair of examples. The first one is more mathematical, while in the second we connect our general settings with a truncated version of the Swanson model, \cite{swan}, which is very well known among the PT-quantum mechanical community, being one non trivial example of manifestly non self-adjoint Hamiltonian which is isospectral to the standard (i.e., self-adjoint) harmonic oscillator.

\subsection{An example with $N=4$}

Let $a$ and $b$ the following four-by-four matrices:
$$
a=\frac{1}{1+\alpha^3}\left(
                        \begin{array}{cccc}
                          (1-\sqrt{2})\alpha^2 & 1+\sqrt{2}\alpha^3 & (\sqrt{2}-1)\alpha & 0 \\
                          -\sqrt{2}\alpha & \sqrt{2}\alpha^2 & \sqrt{2} & \sqrt{3}\alpha(1+\alpha^3) \\
                          \alpha^3 & \alpha & -\alpha^2 & \sqrt{3}(1+\alpha^3) \\
                          0 & 0 & 0 & 0 \\
                        \end{array}
                      \right)
$$
and
$$
b=\frac{1}{1+\alpha^3}\left(
                        \begin{array}{cccc}
                          \alpha & -\alpha^2 & \alpha^3 & 0 \\
                          1+\sqrt{2}\alpha^3 & (\sqrt{2}-1)\alpha & (1-\sqrt{2})\alpha^2 & 0 \\
                          \sqrt{2}\alpha^2 & \sqrt{2} & -\sqrt{2}\alpha & 0 \\
                          -\sqrt{3}\alpha & \sqrt{3}\alpha^2 & \sqrt{3} & 0 \\
                        \end{array}
                      \right),
$$
where $\alpha$ is a real constant different from $-1$. These operators satisfy the commutation rule $[a,b]=\1-4k$, where $k$ is the diagonal matrix on $\Hil_4$ with three zeros and a single one in the main diagonal: $k=diag(0,0,0,1)$. Hence $ka=bk=0$. The hamiltonian $h=ba+\frac{1}{2}(\1-4k)$ looks like
$$
h=\frac{1}{1+\alpha^3}\left(
                        \begin{array}{cccc}
                          \frac{1}{2}(1+3\alpha^2) & \alpha & -\alpha^2 & 0 \\
                          -\alpha^2 & \frac{1}{2}(3+5\alpha^3) & \alpha & 0 \\
                          -2\alpha & 2\alpha^2 & \frac{1}{2}(5+\alpha^3) & 0 \\
                          0 & 0 & 0 & \frac{3}{2} \\
                        \end{array}
                      \right),
$$
which is manifestly not self-adjoint if $\alpha\neq0$: $h\neq h^\dagger$. We will show now how the procedure proposed in this paper can be applied and produces two biorthogonal bases of eigenvectors of $h$ and $h^\dagger$, which are also eigenvectors of the operator $k$. We start by looking at the vector $\varphi_{\frac{3}{2},1}$ which is annihilated by $b$: $b\varphi_{\frac{3}{2},1}=0$. This forces $\varphi_{\frac{3}{2},1}^T$, the transpose of $\varphi_{\frac{3}{2},1}$, to be of the following form:
$
\varphi_{\frac{3}{2},1}^T=\left(
                          \begin{array}{cccc}
                            0 &0 &0 &
                            \gamma_4 \\
                          \end{array}
                        \right),
$
where $\gamma_4$ could be any non zero complex number, (almost) fixed later by the normalization. 

Now, since $a\varphi_{\frac{3}{2},1}=\nu_{\frac{3}{2},1}\varphi_{\frac{5}{2},0}$, we easily check that
$
\varphi_{\frac{5}{2},0}^T=\frac{\sqrt{3}\,\gamma_4}{\nu_{\frac{3}{2},1}}\left(
\begin{array}{cccc}
0 &\alpha &1 &0 \\
                          \end{array}
                        \right).
$
Still, since $a\varphi_{\frac{5}{2},0}=\nu_{\frac{5}{2},0}\varphi_{\frac{3}{2},0}$, we deduce that
$
\varphi_{\frac{3}{2},0}^T=\frac{\sqrt{6}\,\gamma_4}{\nu_{\frac{3}{2},1}\,\nu_{\frac{5}{2},0}}\left(
                          \begin{array}{cccc}
                          0 &1 &0 &0 \\
                          \end{array}
                        \right),
$
while, using the equality $a\varphi_{\frac{3}{2},0}=\nu_{\frac{3}{2},0}\varphi_{\frac{1}{2},0}$, we get
$
\varphi_{\frac{1}{2},0}^T=\frac{\sqrt{6}\,\gamma_4}{\nu_{\frac{3}{2},1}\,\nu_{\frac{5}{2},0}\,\nu_{\frac{3}{2},0}}\left(
                          \begin{array}{cccc}
                          1 &0 &\alpha &0 \\
                          \end{array}
                        \right).
$
An explicit check shows that $a\varphi_{\frac{1}{2},0}=0$, and that $b\varphi_{\frac{5}{2},0}=3\nu_{\frac{3}{2},1}^{-1}\varphi_{\frac{3}{2},1}$, $b\varphi_{\frac{3}{2},0}=2\nu_{\frac{5}{2},0}^{-1}\varphi_{\frac{5}{2},0}$ and $b\varphi_{\frac{1}{2},0}=\nu_{\frac{3}{2},0}^{-1}\varphi_{\frac{3}{2},0}$, as they should.

\vspace{2mm}

The biorthogonal set $\F_\psi^{(4)}$ can be constructed in a similar way: we start looking for a four dimensional vector $\psi_{\frac{3}{2},1}$ which is annihilated by $B=a^\dagger$. This vector is
$
\psi_{\frac{3}{2},1}^T=\left(
                          \begin{array}{cccc}
                          0 &0 &0 &
                          \gamma_4' \\
                          \end{array}
                        \right).
$
Then, acting with $A$ several times on $\psi_{\frac{3}{2},1}$, we find the other vectors of $\F_\psi^{(4)}$. In particular, since $A\psi_{\frac{3}{2},1}=\mu_{\frac{3}{2},1}\psi_{\frac{5}{2},0}$, we find
$
\psi_{\frac{5}{2},0}^T=\frac{\sqrt{3}\,\gamma_4'}{(1+\alpha^3)\mu_{\frac{3}{2},1}}\left(
                          \begin{array}{cccc}
                         -\alpha &\alpha^2 &1 &
                          0 \\
                          \end{array}
                       \right),
$
while $\psi_{\frac{3}{2},0}$ and $\psi_{\frac{1}{2},0}$ can be deduced by acting one or two times with $A$ on $\psi_{\frac{5}{2},0}$. We get
$$
\psi_{\frac{3}{2},0}=\frac{\sqrt{6}\,\gamma_4'}{(1+\alpha^3)\mu_{\frac{3}{2},1}\mu_{\frac{5}{2},0}}\left(
                          \begin{array}{c}
                            \alpha^2 \\
                            1 \\
                            -\alpha \\
                            0 \\
                          \end{array}
                        \right),\qquad \psi_{\frac{1}{2},0}=\frac{\sqrt{6}\,\gamma_4'}{(1+\alpha^3)
                        \mu_{\frac{3}{2},1}\mu_{\frac{5}{2},0}\mu_{\frac{3}{2},0}}\left(
                          \begin{array}{c}
                            1 \\
                            -\alpha \\
                            \alpha^2 \\
                            0 \\
                          \end{array}
                        \right).
$$
A direct check shows that $\left<\varphi_{h',k'},\psi_{h'',k''}\right>=0$ if $(h',k')\neq(h'',k'')$. As for the normalization, the situation is the following: $\left<\varphi_{\frac{3}{2},1},\psi_{\frac{3}{2},1}\right>=1$ if $\gamma_4\gamma_4'=1$; then, $\left<\varphi_{\frac{5}{2},0},\psi_{\frac{5}{2},0}\right>=1$ if $\nu_{\frac{3}{2},1}\mu_{\frac{3}{2},1}=3$; now, $\left<\varphi_{\frac{3}{2},0},\psi_{\frac{3}{2},0}\right>=1$ if $\nu_{\frac{5}{2},0}\mu_{\frac{5}{2},0}=2$. If all these equalities hold, then $\left<\varphi_{\frac{1}{2},0},\psi_{\frac{1}{2},0}\right>=1$ if $\nu_{\frac{3}{2},0}\mu_{\frac{3}{2},0}=1$.  The sets $\F_\varphi^{(4)}$ and $\F_\psi^{(4)}$ obtained in this way are eigenstates of $h$, $h^\dagger$ and of $k$, with the {\em right eigenvalues}.

If we now, for instance, fix $\gamma_4=1$, $\nu_{\frac{3}{2},1}=\mu_{\frac{3}{2},1}=\sqrt{3}$, $\nu_{\frac{5}{2},0}=\mu_{\frac{5}{2},0}=\sqrt{2}$ and $\nu_{\frac{3}{2},0}=\mu_{\frac{3}{2},0}=1$, we obtain the following families of vectors:
$$
\varphi_{\frac{1}{2},0}=\left(
                          \begin{array}{c}
                            1 \\
                            0 \\
                            \alpha \\
                            0 \\
                          \end{array}
                        \right),\quad \varphi_{\frac{3}{2},0}=\left(
                          \begin{array}{c}
                            \alpha \\
                            1 \\
                            0 \\
                            0 \\
                          \end{array}
                        \right),\quad \varphi_{\frac{5}{2},0}=\left(
                          \begin{array}{c}
                            0 \\
                            \alpha \\
                            1 \\
                            0 \\
                          \end{array}
                        \right),\quad \varphi_{\frac{3}{2},1}=\left(
                          \begin{array}{c}
                            0 \\
                            0 \\
                            0 \\
                            1 \\
                          \end{array}
                        \right),
$$
while
$$
\psi_{\frac{1}{2},0}=\frac{1}{(1+\alpha^3)}\left(
                          \begin{array}{c}
                            1 \\
                            -\alpha \\
                            \alpha^2 \\
                            0 \\
                          \end{array}
                        \right),\, \psi_{\frac{3}{2},0}=\frac{1}{(1+\alpha^3)}\left(
                          \begin{array}{c}
                            \alpha^2 \\
                            1 \\
                            -\alpha \\
                            0 \\
                          \end{array}
                        \right), \, \psi_{\frac{5}{2},0}=\frac{1}{(1+\alpha^3)}\left(
                          \begin{array}{c}
                            -\alpha \\
                            \alpha^2 \\
                            1 \\
                            0 \\
                          \end{array}
                        \right), \, \psi_{\frac{3}{2},1}=\left(
                          \begin{array}{c}
                            0 \\
                            0 \\
                            0 \\
                            1 \\
                          \end{array}
                        \right).
$$
If we fix, for concreteness, $\alpha={1}{2}$, we find in particular that the operators $S_\varphi$ and $S_\psi$ are
$$
S_\varphi=\frac{1}{4}\left(
\begin{array}{cccc}
 5 & 2 & 2 & 0 \\
 2 & 5 & 2 & 0 \\
 2 & 2 & 5 & 0 \\
 0 & 0 & 0 & 4 \\
\end{array}
\right),
\qquad S_\psi=\frac{1}{27}\left(
\begin{array}{cccc}
 28 & -8 & -8 & 0 \\
 -8 & 28 & -8 & 0 \\
 -8 & -8 & 28 & 0 \\
 0 & 0 & 0 & 27 \\
\end{array}
\right).
$$
Following Section \ref{sectIII.1}, and (\ref{38}) in particular, we get the following {\em self-adjoint version} of the system:
%$$
%S_\varphi^{1/2}=\left(
%\begin{array}{cccc}
% \frac{1}{2}+\frac{1}{\sqrt{3}} & \frac{1}{6} \left(3-\sqrt{3}\right) & \frac{1}{6} \left(3-\sqrt{3}\right) & 0 \\
% \frac{1}{6} \left(3-\sqrt{3}\right) & \frac{1}{2}+\frac{1}{\sqrt{3}} & \frac{1}{6} \left(3-\sqrt{3}\right) & 0 \\
% \frac{1}{6} \left(3-\sqrt{3}\right) & \frac{1}{6} \left(3-\sqrt{3}\right) & \frac{1}{2}+\frac{1}{\sqrt{3}} & 0 \\
% 0 & 0 & 0 & 1 \\
%\end{array}
%\right)
%$$
%and
%$$
%S_\psi^{1/2}=\left(
%\begin{array}{cccc}
% \frac{2}{9} \left(1+2 \sqrt{3}\right) & \frac{-2}{9} \left(-1+\sqrt{3}\right) & \frac{-2}{9} \left(-1+\sqrt{3}\right) & 0 \\
% \frac{-2}{9}  \left(-1+\sqrt{3}\right) & \frac{2}{9} \left(1+2 \sqrt{3}\right) & \frac{-2}{9}  \left(-1+\sqrt{3}\right) & 0 \\
% \frac{-2}{9}  \left(-1+\sqrt{3}\right) & \frac{-2}{9}  \left(-1+\sqrt{3}\right) & \frac{2}{9} \left(1+2 \sqrt{3}\right) & 0 \\
% 0 & 0 & 0 & 1 \\
%\end{array}
%\right).
%$$
%We can now use these operators as in (\ref{38}) to compute the {\em self-adjoint version} of the model. In particular we get, using (\ref{38}),
$$
c=\left(
\begin{array}{cccc}
 \frac{1}{9} \left(1+\sqrt{2}+\sqrt{3}-\sqrt{6}\right) & \frac{1}{9} \left(4+\sqrt{2}+2 \sqrt{3}\right) & \frac{1}{9} \left(-2+\sqrt{2}+\sqrt{6}\right) & \frac{1}{\sqrt{3}}-1 \\
 \frac{1}{9} \left(1-2 \sqrt{2}-\sqrt{3}\right) & \frac{1}{9} \left(-2+\sqrt{2}+\sqrt{6}\right) & \frac{2}{9} \left(2+2 \sqrt{2}-\sqrt{3}+\sqrt{6}\right) & \frac{1}{\sqrt{3}} \\
 \frac{1}{9} \left(1+4 \sqrt{2}-2 \sqrt{6}\right) & \frac{1}{9} \left(1+\sqrt{2}+\sqrt{3}-\sqrt{6}\right) & \frac{1}{9} \left(1-2 \sqrt{2}-\sqrt{3}\right) & 1+\frac{1}{\sqrt{3}} \\
 0 & 0 & 0 & 0 \\
\end{array}
\right),
$$
and the following o.n. basis:
$$
e_{\frac{1}{2},0}=\left(
\begin{array}{c}
 \frac{1}{3}+\frac{1}{\sqrt{3}} \\
 \frac{1}{3}-\frac{1}{\sqrt{3}} \\
 \frac{1}{3} \\
 0 \\
\end{array}
\right),\quad e_{\frac{3}{2},0}=\left(
\begin{array}{c}
 \frac{1}{3} \\
 \frac{1}{3}+\frac{1}{\sqrt{3}} \\
 \frac{1}{3}-\frac{1}{\sqrt{3}} \\
 0 \\
\end{array}
\right), \quad e_{\frac{5}{2},0}=\left(
\begin{array}{c}
 \frac{1}{3}-\frac{1}{\sqrt{3}} \\
 \frac{1}{3} \\
 \frac{1}{3}+\frac{1}{\sqrt{3}} \\
 0 \\
\end{array}
\right), \quad e_{\frac{3}{2},1}=\left(
\begin{array}{c}
 0 \\
 0 \\
 0 \\
 1 \\
\end{array}
\right).
$$
It is easy to check that $c$, together with its adjoint $c^\dagger$, satisfies (\ref{11}), and that the vectors $e_{h',k'}$ are eigenstates of $H_0=c^\dagger c+\frac{1}{2}(\1-4K)$ and $K$, with the {\em right} eigenvalues. So our original system can be mapped into a system as those described in \cite{buc}. However, in order to perform such a mapping, it should be stressed that the eigenvectors $\varphi_{h',k'}$ (or the $\psi_{h',k'}$) should be found first, to construct $S_\varphi$ and its inverse, $S_\psi$, and then their square roots. These are in fact the essential ingredients of formula (\ref{38}).

\subsection{A truncated Swanson model}

In \cite{buc} the commutation rule (\ref{11}) have been used to consider a truncated version of the harmonic oscillator, living in the finite dimensional Hilbert space $\Hil_N$, and then considering its limit for diverging $N$. In \cite{bgv} it has been discussed that the self-adjoint Hamiltonian of the oscillator produces, using similarity transformations, several non self-adjoint quadratic Hamiltonians with known spectra and eigenstates which may, or may not, form bases for the infinite-dimensional Hilbert space where the model is defined. Among these Hamiltonians, one can recover the ones for the shifted harmonic oscillator and for the Swanson model. All these systems can be described in terms of pseudo-bosonic operators. Hence it might be interesting to consider the truncated versions of this models. For instance, following \cite{baginbagbook}, we introduce
$$
H_\theta=\frac{1}{2}\left(p^2+x^2\right)-\frac{i}{2}\,\tan(2\theta)\left(p^2-x^2\right), $$ where $\theta$ is a real parameter taking value in
$I=\left(-\frac{\pi}{4},\frac{\pi}{4}\right)\setminus\{0\}$.  We assume here that $[x,p]=i(\1-Nk)$. Hence $H_\theta$ can be rewritten as
$ H_\theta=\omega_\theta\left(B_\theta\,A_\theta+\frac{1}{2}(\1-Nk)\right), $ where $\omega_\theta=\frac{1}{\cos(2\theta)}$ is well
defined because $\cos(2\theta)\neq0$ for all $\theta\in I$, and where the operators $A_\theta$ and $B_\theta$ are defined as in \cite{baginbagbook}:
$$\left\{
\begin{array}{ll}A_\theta=\frac{1}{\sqrt{2}}\left(e^{i\theta}x+ie^{-i\theta}p\right),\\
B_\theta=\frac{1}{\sqrt{2}}\left(e^{i\theta}x-ie^{-i\theta}p\right),\end{array} \right. $$
which satisfy the commutation rule $[A_\theta,B_\theta]=\1-Nk$. Of course, if $\theta\neq0$, $H_\theta\neq H_\theta^\dagger$. The $N$ eigenstates of $H_\theta$ and  those of $H_\theta^\dagger$ can be constructed as shown previously. In particular they are also eigenstates of the projection operator $k$, and correspond to the eigenvalues $\frac{\omega_\theta}{2}$, $\frac{3\,\omega_\theta}{2}$, $\frac{5\,\omega_\theta}{2}$ and so on, depending on the explicit value of $N$. Of course, as in the infinite dimensional Swanson model, since $H_\theta$ and $H_\theta^\dagger$ have the same eigenvalues, we can imagine that they satisfy a suitable intertwining relations, and in fact we could explicitly deduce that
$$
H_\theta S_\varphi=S_\varphi H_\theta^\dagger.
$$
Notice that this is much better than we get for the infinite dimensional case, \cite{baginbagbook}. In that case, in fact, we are only able to prove that the analogous of the intertwining equation in above holds on each eigenstate of $H_\theta^\dagger$, but the set of these vectors is not a basis for the Hilbert space: it is only a $\D$-quasi basis. Here this problem does not exist since in the present situation the set of eigenstates of $H_\theta$ is surely a basis for the finite-dimensional space $\Hil_N$. So, in a sense, the truncated Swanson model is {\em better} than the original one. 

Of course, it is clear that this {\em truncation} works also for many other quantum mechanical systems which are similar to the harmonic oscillator, as those listed in \cite{bgv}. In particular, for instance, it can be applied to the shifted harmonic oscillator whose Hamiltonian, $H_\beta=\frac{\beta}{2}\left(p^2+x^2\right)+i\sqrt{2}\,p$,  $\beta>0$, is manifestly non self-adjoint. In principle, then, we can expect that this kind of truncation can be relevant is applications, at least for some particular systems. For instance, a similar framework was used in connection with a biological system in \cite{bagentropy}. What we like in the operators used here is that they obey interesting, and simple, commutation relations so that they can produce more relevant mathematical and physical results.

\section{Conclusions}\label{sectV}

We have proposed a deformed version of the truncated harmonic oscillator which, in our opinion, is particularly interesting in connection with $PT$-quantum mechanics and with its relatives. We have shown how two biorthogonal bases can be constructed, using raising and lowering operators which are not necessarily related by an adjoint operation, and that these bases are eigenstates of two Hamiltonians, one the adjoint of the other, connected by suitable intertwining relations. In particular, these results extend, and improve, those found in \cite{abg}, where two biorthogonal bases were used to define two different pairs of ladder operators but where no closed commutation rule was deduced.

We have also discussed in details an example in $N=4$, and a possible application to the Swanson model, and in particular we have deduced that the basis property for its truncated, non-self-adjoint, version is satisfied.

\section*{Acknowledgements}

FB gratefully thanks B. Bagchi for useful discussions and for suggesting the reading of \cite{buc} and \cite{bagchi}. He also acknowledges support from the GNFM of Indam and from the University of Palermo.

\end{document}